\documentclass[lettersize,journal]{IEEEtran}
\usepackage{amsmath,amssymb,amsfonts}
\usepackage{algorithm}
\usepackage{algpseudocode}
\usepackage{array}
\usepackage{textcomp}
\usepackage{stfloats}
\usepackage{url}
\usepackage{verbatim}
\usepackage{graphicx}
\usepackage{cite}
\usepackage{multicol}

\usepackage{subcaption}
\usepackage{booktabs}
\newcommand\mytab[1]{\begin{tabular}[t]{@{}c@{}} #1 \end{tabular}}
\newcommand\mc[2]{\multicolumn{#1}{c}{#2}}
\usepackage{siunitx}
\usepackage{textcomp}

\usepackage{hyperref}
\hypersetup{
    colorlinks=true,
    allcolors=black,
    pdftitle={HOOD: Real-Time Robust Human Presence and Out-of-Distribution Detection with Low-Cost FMCW Radar},
    pdfpagemode=FullScreen,
    }

\begin{document}
\title{HOOD: Real-Time Human Presence and Out-of-Distribution Detection Using FMCW Radar
}
\author{Sabri Mustafa Kahya, Muhammet Sami Yavuz, and Eckehard Steinbach

\thanks{Sabri Mustafa Kahya, Muhammet Sami Yavuz, and  Eckehard Steinbach are with the Technical University of Munich, School of Computation, Information and Technology, Department of Computer Engineering, Chair of Media Technology, Munich Institute of Robotics and Machine Intelligence (MIRMI) (e-mail: \{mustafa.kahya, sami.yavuz, eckehard.steinbach\}@tum.de). }
\thanks{Muhammet Sami Yavuz is also with Infineon Technologies AG.}

}



\maketitle

\begin{abstract}
Detecting human presence indoors with millimeter-wave frequency-modulated continuous-wave (FMCW) radar faces challenges from both moving and stationary clutter. This work proposes a robust and real-time capable human presence and out-of-distribution (OOD) detection method using 60 GHz short-range FMCW radar. HOOD solves the human presence and OOD detection problems simultaneously in a single pipeline. Our solution relies on a reconstruction-based architecture and works with radar macro and micro range-Doppler images (RDIs). HOOD aims to accurately detect the presence of humans in the presence or absence of moving and stationary disturbers. Since HOOD is also an OOD detector, it aims to detect moving or stationary clutters as OOD in humans' absence and predicts the current scene's output as ``\textit{no presence}." HOOD performs well in diverse scenarios, demonstrating its effectiveness across different human activities and situations. On our dataset collected with a 60 GHz short-range FMCW radar, we achieve an average AUROC of 94.36\%. Additionally, our extensive evaluations and experiments demonstrate that HOOD outperforms state-of-the-art (SOTA) OOD detection methods in terms of common OOD detection metrics. Importantly, HOOD also perfectly fits on Raspberry Pi 3B+  with an ARM Cortex-A53 CPU, which showcases its versatility across different hardware environments. Videos of our human presence detection experiments are available at: \url{https://muskahya.github.io/HOOD}. 
\end{abstract}

\begin{IEEEkeywords} Human presence detection, out-of-distribution detection, 60 GHz FMCW radar, deep neural networks \end{IEEEkeywords}

\maketitle

\section{Introduction}
\label{sec:introduction}

\IEEEPARstart{I}{n} recent years, radars have gained huge attention among other types of sensors due to their robustness against environmental conditions such as lighting, smoke, and rain, as well as their privacy-preserving nature. Therefore they are utilized for several applications such as reliable detection and interpretation of human gestures \cite{gesture_recog, hand_gest,hand_gest2,hand_gest_IoT}, extracting vital signs like heart rate and respiration rate \cite{vital_sign,vital_sign2,heartbeat_est}, accurate human activity classification \cite{human_activity,human_activity2,human_activity3}, and real-time people counting in various environments\cite{people_counting, people_counting2, people_counting3}. These applications cover multiple areas, particularly in human-computer interaction (HCI) and the Internet of Things (IoT). Here, they are crucial for improving smart home automation, healthcare monitoring, and other IoT solutions. Incorporating radar technology into IoT applications is key to creating smarter environments, boosting both efficiency and safety. Notably, detecting human presence is an important application in this context that takes advantage of radar's distinctive abilities.

Real-time and robust human presence detection solutions using radars can enhance safety, energy efficiency, and personalized experiences. For instance, FMCW radar-based human presence detection systems can detect threats and unauthorized access indoors, enhancing security in restricted areas. Integration in smart homes enhances personalized experiences and improves energy efficiency by automatically adjusting environmental settings like lighting and heating based on accurate human presence detection.

In addition, human presence detection utilizing short-range
radar sensors presents its own challenges. In real-world scenarios,
both indoor and outdoor environments, there are many moving
and stationary disturbances. With these disturbances present,
extracting the signals coming from humans becomes challenging. In non-human environments, disturbers' macro or micro movements may produce human-like signals and create false detections for human presence detection applications. In our previous work \cite{MCROOD}, we addressed this issue and proposed an OOD detection-based solution which can identify and differentiate between a disturber and a human (sitting, standing, or walking) only when either of them is present in front of the radar sensor. However, when both disturbers and humans are in front of the sensor, the reflected signals from the disturbers may interfere with those from humans. The previous work mentioned was incapable of solving this issue and tended to classify the scenario as OOD due to the signal dominance of disturbers. This work addresses every edge case and proposes a robust presence detection solution. Similarly, this work approaches the issue as an OOD detection problem.

OOD detection has gained significant attention due to its crucial role in ensuring the secure and reliable deployment of modern deep learning (DL) architectures. Despite the impressive performance of DL models in various applications, they often make closed-world assumptions, which are unrealistic in real-world scenarios. This can result in the misclassification of OOD samples, potentially leading to severe consequences in safety-critical domains like healthcare and autonomous driving. To address this challenge, researchers have developed various OOD detection strategies \cite{b25,b26,b27,b29,b24}, primarily focused on image or video data. In this paper, we propose a reconstruction-based approach by utilizing radar macro and micro range-Doppler images (RDIs) for improved OOD detection in the radar domain. 

\begin{figure*}[htbp]
\vspace{-0.6cm}
\centerline{\includegraphics[width=0.87\linewidth]{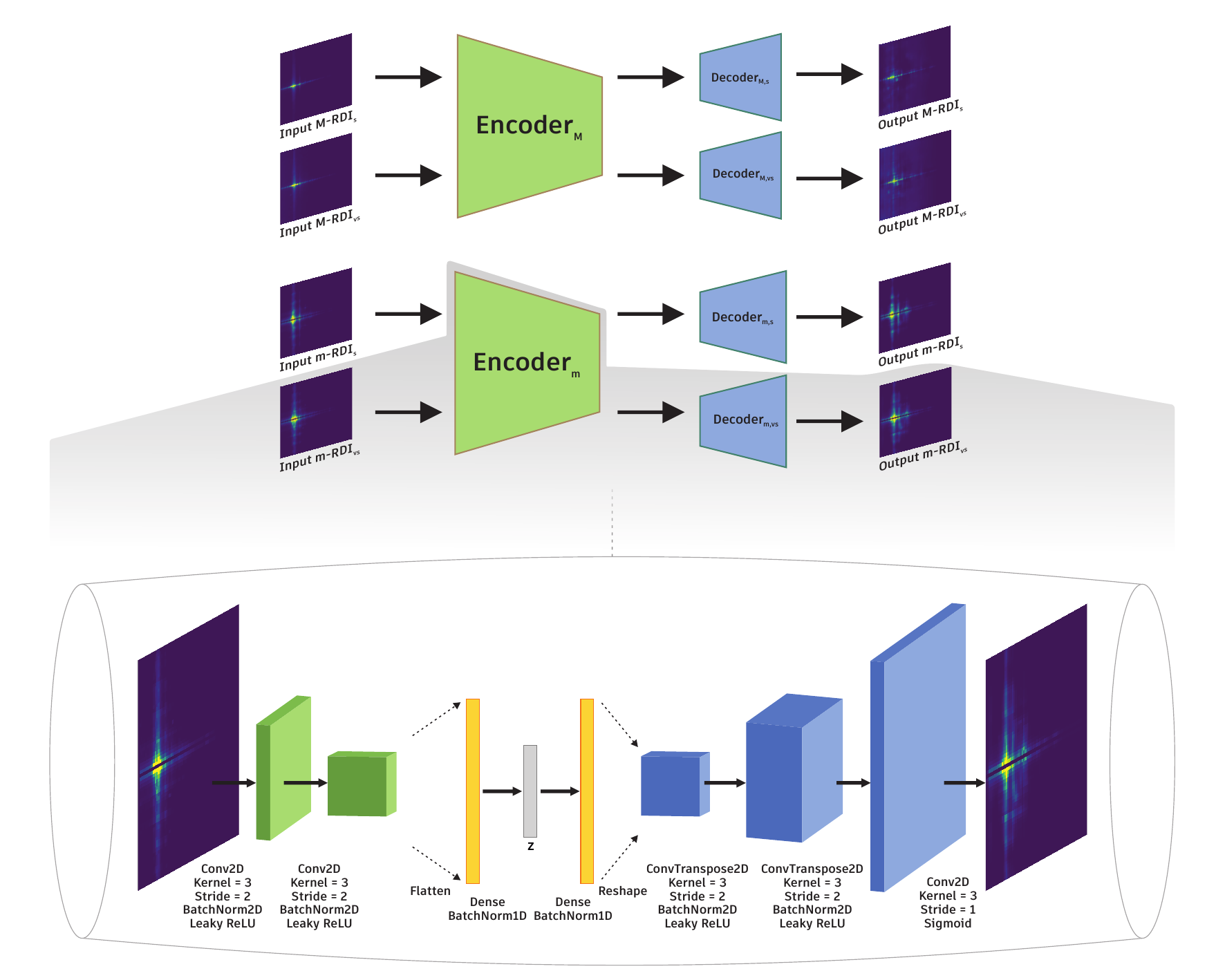}}
\caption{This figure presents the high-level structure of HOOD (top) and provides a detailed view of one encoder-decoder pair within the architecture (bottom). Here the input data are radar macro and micro RDIs from static and very-static categories. The RDIs are processed through the network to extract meaningful representations. The architecture follows a reconstruction-based approach to reconstruct and interpret the input RDIs. In the high-level structure, the top and bottom encoders share the same architecture and are responsible for encoding macro and micro RDIs, respectively.  The four decoders, in top-to-bottom order, decode macro RDIs for static activities, macro RDIs for very-static activities, micro RDIs for static activities, and micro RDIs for very-static activities, also having identical structures.}
\label{fig:pipline}
\end{figure*}

OOD methods typically do not incorporate OOD samples during training. Therefore, their pipelines are never exposed to samples outside the training distribution. Moreover, these methods do not evaluate their performance using samples that simultaneously contain both in-distribution (ID) and OOD classes. In our work, in addition to samples containing only IDs, we also leverage samples that consist of both IDs and OODs (ID + OOD is also ID in our case). This approach enables us to detect IDs even in the presence of OODs, aiming to enhance the overall detection capability. This paper introduces a new DL architecture (see Figure \ref{fig:pipline}) designed for human and OOD detection.
Additionally, we incorporate an enhanced RESPD \cite{MCROOD} (E-RESPD) technique as a pre-processing method to detect tiny and almost invisible human body movements in addition to respiration movement, thereby facilitating the human and OOD detection process. With the utilization of E-RESPD, the detection of humans remains possible even in cases where individuals intentionally hold their breath. HOOD relies on a reconstruction-based architecture that comprises a multi-encoder multi-decoder system and E-RESPD. Generally, OOD detectors employ simple thresholding for detection, where a threshold is defined based on a scoring function to classify samples as either ID or OOD. HOOD utilizes multi-thresholding to detect humans in the presence or absence of OODs and to detect OODs themselves. We divide the human detection process into two categories, namely static human(s) like standing and very-static human(s) like sitting and lying. Dynamic human movement walking has a unique and dominant signature that is easily detectable in the presence or absence of disturbers using conventional signal processing or learning-based approaches. Therefore, we do not include the walking scenario in this work. For this, we refer to our previous work that shows superior performance on the walking scenario \cite{MCROOD}. Our key contributions are as follows:

\begin{itemize}
    \item We propose a novel human presence detector called HOOD, which utilizes a reconstruction-based architecture and operates on radar macro and micro RDIs. The architecture is composed of a multi-encoder multi-decoder system. Each encoder-decoder pair collaboratively focuses on detecting static and very-static human(s) with the assistance of micro RDIs. Additionally, for scenarios where humans exhibit macro movements like hand waving when they are static or very-static, HOOD employs macro RDIs for detection. HOOD is a versatile human detector that does not rely on predefined classes and can detect humans engaged in various activities.
    
    \item We propose HOOD as an effective OOD detector in addition to its role as a human presence detector. HOOD can identify both moving and stationary clutter in scenarios where no human presence is detected, classifying them as OOD samples.
    
    \item We introduce E-RESPD as an improved iteration of RESPD. By integrating E-RESPD, we achieve enhanced detection of human respiration movement. It can also accurately capture minute and nearly invisible body movements, even when individuals intentionally refrain from breathing. This concept improves the overall classification performance.

    \item We achieve promising results in terms of AUROC, AUPR, and FPR95. Also, our extensive evaluations on our dataset highlight the superiority of HOOD over SOTA methods in terms of common metrics.
 
\end{itemize}
 In the following section, we discuss relevant previous work. Section \ref{section:system} explains radar system design and pre-processing including E-RESPD. Section \ref{section:method} delves into the details of our proposed method, outlining its key components and methodologies. Section \ref{section:experiment} presents the experiments conducted to validate our approach, including the experimental setup, data collection, and analysis. Finally, Section \ref{section:conclusion} draws conclusions and suggests future research directions.

\section{Related Work}
\label{section:related_work}
In the literature, there are several radar-based human presence detection works \cite{human_presence, human_presence2}. These works mostly utilize conventional signal processing approaches. \cite{human_presence3} described a real-time presence detection pipeline using $\SI{24}{\giga\hertz}$ FMCW radar in an office environment. The method aims to detect tiny human body movements like typing. However, it does not evaluate complete static human scenarios. In \cite{occupancy_sens}, the authors proposed an occupancy sensing application with $\SI{60}{\giga\hertz}$ FMCW radar for heating, ventilation, and air-conditioning (HVAC) systems. They created a novel pipeline to detect vital, micro, and macro Doppler movements from various human activities. In \cite{breath_rate}, the proposed solution aims to detect the presence only by minute movements of breathing. They tested their pipeline only with sleeping babies and sitting adults. The studies mentioned earlier aim to identify the presence of people using unique methods. However, they overlook both moving and stationary clutter, which is common in indoor and outdoor environments in the real world. \cite{human_presence4} described a presence detection method using a UWB radar by considering outdoor disturbers. The algorithm uses the human body's scattered UWB waveforms to detect the human target(s) in urban environments. However, its classical likelihood ratios-based detection pipeline requires dynamic human motion, and it is tested only with cars as outdoor disturbers. In the follow-up work \cite{human_presence5}, the authors developed an improved pipeline to simultaneously detect and track multiple human and non-human targets by only considering walking person activity. In their experiments, they only use an aluminum foil-covered basketball representing a small moving animal as a non-human object. For detecting standing or lying individuals in the presence of clutters, Kim et al. developed a two-stage detection algorithm \cite{human_presence6} using a UWB radar. It includes initial detection in the range domain and subsequent detection in the frequency domain.
Similarly, for a human presence detection application, \cite{human_presence7} proposed a novel moving and stationary clutter suppression technique to identify the characteristic respiratory signal better. Our work is an activity-free human presence detector and detects human(s) in the presence or absence of various disturbers. Different from the literature, we approach the human presence detection task as an OOD detection problem.

The field of OOD detection is relatively new. A foundational work by Hendrycks \& Gimpel introduces the baseline method \cite{b1}, which relies on maximum softmax probabilities (MSP). Their main assertion is that deep neural networks assign higher confidence values to ID samples compared to OOD samples during inference. As a result, a simple thresholding technique can be used to differentiate between ID and OOD samples. The ODIN approach, presented in \cite{b2}, enhances \cite{b1} by incorporating temperature scaling and input perturbations to amplify the softmax scores of ID samples. This approach is further extended by introducing a model ensembling technique \cite{b3}. In a different study, Liu et al. \cite{b7} proposed an energy-based method for OOD detection, employing the $LogSumExp$ function on the logit layer and noting that ID samples tend to have lower energy scores than OOD samples. The Mahalanobis distance-based detector (MAHA) \cite{b4} is designed to identify OOD samples by leveraging intermediate feature representations of deep neural networks. MAHA constructs a class conditional Gaussian distribution for each ID class and evaluates the Mahalanobis distance between a test sample and each class conditional Gaussian distribution to calculate an OOD score. Additionally, MAHA incorporates layer ensembling and input perturbation techniques.

Similarly, \cite{b5, simple_maha} employ intermediate representations for OOD detection. \cite{hendrycks2022scaling} introduced two different methods relying on maximum logit and minimum KL divergence information. ReAct \cite{b28} addresses overconfident predictions by employing activation truncation on the penultimate layer, whereas DICE \cite{dice} uses weight sparsification for OOD detection. In \cite{b6}, the authors introduced uniform noise and argued that OOD samples resemble noise more closely than ID samples. To estimate the OODs using the distance information, \cite{b31} adopts a deep-nearest-neighbor approach. While the aforementioned methods are compatible with any pre-trained model, some require OOD samples for hyperparameter tuning. G-ODIN \cite{b30} introduced an enhanced training scheme built upon \cite{b2} to achieve improved detection capabilities.
Hendrycks et al. \cite{b8} proposed the Outlier Exposure (OE) technique, which involves integrating a limited number of OOD samples during the fine-tuning or training process. They introduced a novel loss function that promotes the convergence of softmax probabilities for OOD samples towards a uniform distribution. OECC \cite{b10} follows a similar approach with the inclusion of a novel loss function consisting of two distinct regularization terms. In \cite{b11}, a GAN architecture is utilized, along with a specialized loss function, to generate synthetic OOD samples for OE. Moreover, other studies \cite{b9, poem} in the literature also benefit from OOD samples to provide more powerful OOD detectors.
The significance of gradient information in the OOD detection task is highlighted in some studies \cite{b2, b14}. GradNorm \cite{b14} leverages the magnitudes of gradients to differentiate between ID and OOD samples. To compute the gradients, the KL divergence between the softmax output and a uniform distribution is backpropagated. According to this approach, ID samples exhibit higher gradient magnitudes compared to OOD samples. 

In a reconstruction-based method \cite{b15} that utilizes autoencoders, a combination of the reconstruction loss and Mahalanobis distance in the latent code is used as an OOD score. Other studies, such as \cite{b16,b17}, utilize Variational Autoencoders (VAEs) and their latent representations for OOD detection. Although most OOD detection methods rely on convolutional neural network (CNN) architectures, there are also notable alternatives, such as the approaches mentioned in \cite{b18, b19, b20}, which employ pre-trained transformers with attention mechanisms like VIT \cite{b21} and BERT \cite{b22}. These transformer-based methods have achieved SOTA results on various benchmark datasets.

In addition to the previously mentioned studies, several other approaches have been proposed within the radar domain for OOD detection.
In \cite{RB-OOD}, the authors introduced a reconstruction-based OOD detector specifically designed for $\SI{60}{\giga\hertz}$  FMCW radar. They aim to distinguish a walking human from various moving disturbers. The proposed method utilizes two scores, incorporating patch-based reconstruction and energy scores, for OOD detection. \cite{MCROOD} proposed MCROOD as a multi-class radar OOD detector which tries to detect any clutter other than a human sitting, standing, or walking as OOD. It relies on a reconstruction-based one-encoder multi-decoder architecture and a simple pre-possessing idea RESPD. For OOD detection, it uses multi-thresholding. In \cite{b36}, a hand gesture recognition system was introduced that utilizes FMCW radar technology and incorporates the capability to identify OOD input data. In \cite{b23}, a comparison is made between traditional and deep learning techniques for near OOD detection, utilizing synthetically generated low-resolution radar micro-Doppler signatures. Additionally, \cite{b37} presented a meta-reinforcement learning approach that enhances radar tracking reliability and includes support for OOD detection.

\section{RADAR SYSTEM DESIGN}
\label{section:system}
The research presented in this paper utilizes the FMCW radar chipset provided by Infineon, known as BGT60TR13C (The FMCW radar configuration for this study is provided in Table \ref{tab:radar_conf}). This radar chipset, operating at a frequency range of $\SI{57}{\giga\hertz}$ to $\SI{64}{\giga\hertz}$, offers the ability to customize the chirp duration ($T\textsubscript{c}$). The radar chipset has one transmit (Tx) antenna and three receiver (Rx) antennas. The receiver antennas are arranged in an L-shape configuration, and the distance between neighboring receiving antennas is half-wavelength. The gain of each receive and transmit antenna is 10 
\unit[qualifier-mode = combine]
{\deci\bel\of{i}}  and 6 \unit[qualifier-mode = combine]{\deci\bel\of{i}}, respectively.

The chipset follows the fundamental principles of FMCW radar. The reference frequency ($f\textsubscript{ref}  = \SI{80}{\mega\hertz}$) of a voltage-controlled oscillator (VCO) is maintained by a phase-locked loop (PLL) in the transmission path. By modifying the divider value and using a tuning voltage within the range of $\SI{1}{\volt}$ to $\SI{4.5}{\volt}$, the chipset generates precise linear frequency chirps ($N\textsubscript{c}$) from $\SI{57}{\giga\hertz}$ to $\SI{64}{\giga\hertz}$. The transmitted signal (chirp signal) in an FMCW system is given by
\[
s(t) = \exp\left(j2\pi \left(f_c t + \frac{S}{2} t^2\right)\right), \quad \forall 0 < t < T_c
\]
where \(f_c\) is the center frequency, \(S\) is the chirp rate (frequency sweep rate) calculated as the ratio of the signal bandwidth \(B\) to the chirp duration ($S = \frac{B}{T\textsubscript{c} }$). The transmitted signal is backscattered by objects in the environment, resulting in a time delay, i.e.,  round trip delay due to the range of the target from the radar and its velocity.
The receive antennas capture the delayed signal. A mixture of received and transmitted signals undergoes a low-pass filter to acquire the intermediate frequency (IF) signal, which is then used to apply Analog-to-digital converter (ADC) with a sampling frequency of $\SI{2}{\mega\hertz}$ with $12$-bit accuracy. The data is reorganized and molded into a frame referred to as $N\textsubscript{c}  \times N\textsubscript{s}$. Within this frame, each column comprises $N\textsubscript{c}$ samples denoted as slow-time samples, while each row consists of $N\textsubscript{s}$ samples known as fast-time samples. This configuration applies to a total of three receiving antennas. As a result, the digitized signal takes on the shape of $N\textsubscript{Rx} \times N\textsubscript{c} \times N\textsubscript{s} $, making it ready for further signal processing.

\subsection{Pre-processing}

In our study, we utilize both macro and micro RDIs for our architecture. These RDIs are generated by applying pre-processing operations to the digitized signal ($N\textsubscript{Rx} \times N\textsubscript{c} \times N\textsubscript{s} $).

For the extraction of the macro RDI, we initially apply the Range-FFT on the fast time to determine the object's range. Mean removal is also performed to obtain one channel range information out of three Rx channels. The range data is then processed using the Moving Target Identification (MTI) function, which effectively eliminates static targets. Finally, we apply Doppler-FFT along the slow time to capture the phase across chirps, allowing us to obtain the desired macro RDI.

To extract the micro RDI, we begin with the Range-FFT to obtain range information. To capture more detailed information, we stack 8 range spectrograms. Additionally, mean removal is performed in both the fast and slow times to reduce noise. We further enhance the target signals by applying Sinc filtering. Finally, Doppler-FFT along the slow time is utilized to identify target movement in the micro RDI.

\begin{table}[h]
    \caption{\footnotesize FMCW Radar Configuration Parameters }
    \centering
     \footnotesize
    \begin{tabular}{@ {\extracolsep{10pt}} ccc}
    \toprule

    \centering
    Configuration name & Symbol & Value \\
    \midrule
    Number of transmit antennas & $N\textsubscript{Tx}$  & 1  \\
    Number of receive antennas & $N\textsubscript{Rx}$  & 3  \\
    Chirps per frame & $N\textsubscript{c}$ & 64  \\
    Samples per chirp & $N\textsubscript{s}$  & 128 \\
    Frame period & T\textsubscript{f} & 50 \si{\ms}  \\
    Chirp to chirp time & $T\textsubscript{cc}$ & 391.55 \si{\us} \\
    Ramp start frequency & $f\textsubscript{min}$ & $\SI{60.1}{\giga\hertz}$\\ Ramp stop frequency & $f\textsubscript{max}$ &  $\SI{61.1}{\giga\hertz}$\\
    Bandwidth & $B$ & $\SI{1}{\giga\hertz}$\\ 
    \bottomrule
    \end{tabular}

    \label{tab:radar_conf}
\end{table}

\subsection{E-RESPD}

Respiration detector (RESPD) \cite{MCROOD} is a pre-processing technique designed for RDIs. Its objective is to capture the human respiration movement by utilizing 50 consecutive frames, equivalent to 2.5 seconds in the radar configuration. This is accomplished through a sliding-window approach with a window size of 50 frames. Within each window, the frames are summed, and the resulting value is written onto the first frame of the window. The window is then shifted by one frame, and the process is repeated until the end of the recording.

\begin{figure*}[htbp]
\centerline{\includegraphics[width=0.98\linewidth]{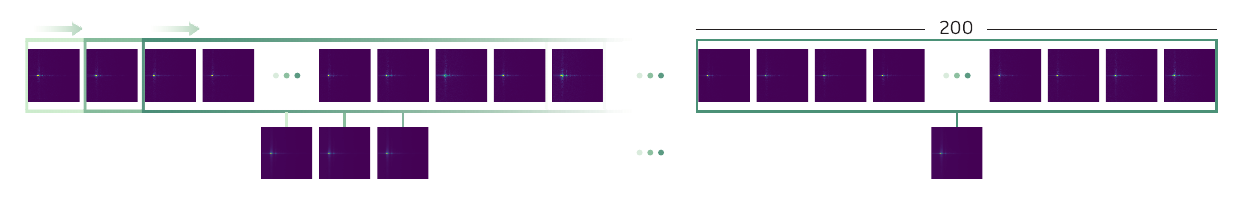}}
\caption{Illustration of the working mechanism of E-RESPD for pre-processing of macro and micro RDIs. E-RESPD utilizes a sliding-window approach. The window size is set to 200 frames, equivalent to 10 seconds of data acquisition from the radar sensor. The process starts with the first 200 frames being summed and written back to the first frame. Then, the window is shifted by one, and the next 200 frames in the window are summed and written back to the second frame. This process continues until the end of the data. }
\label{fig:erespd}
\end{figure*}

The enhanced respiration detector E-RESPD on the other hand operates on both macro and micro RDIs. In E-RESPD the window size is increased from 50 to 200 frames, representing 10 seconds in the radar configuration, enabling the modeling of subtle and nearly imperceptible human body movements. This includes scenarios such as detecting movements when a person intentionally holds his breath. Figure \ref{fig:erespd} provides a detailed illustration of the overall working mechanism of E-RESPD.

\section{Proposed Method}
\label{section:method}

In this section, we provide a detailed explanation of our human and OOD detector, HOOD. HOOD utilizes a reconstruction-based multi-encoder multi-decoder architecture, combining multiple encoders and decoders for robust presence detection. By training on scenarios, including static and very-static human activities with and without disturbers, our model learns to generalize the solution across various cases. We employ a multi-thresholding technique based on reconstruction errors to classify samples as human or OOD. The architecture, training process, and algorithmic implementation of HOOD are presented in detail in the following sections.

\subsection{Architecture and Training}

HOOD relies on a reconstruction-based multi-encoder ($2$) multi-decoder ($4$) architecture. Each encoder shares the same architecture, while each decoder also follows an identical structure. The encoders are responsible for encoding the macro and micro RDIs, respectively. On the other hand, the decoders are responsible for decoding the macro and micro RDIs of the static and very-static categories. The encoder architecture consists of three primary blocks. The initial two blocks involve the sequential application of $2$D convolution, followed by $2$D batch normalization and LeakyReLU activation. In these blocks, we respectively employ $16$ and $64$ filters with a $3 \times 3$ kernel size for the convolutional layers. Downsampling is achieved in those blocks with a stride of $2 \times 2$. The final block incorporates flattening and fully connected (dense) layers, followed by $1$D batch normalization, to generate the latent representation. On the other hand, the decoder architecture comprises four primary blocks. In the first block, the latent code undergoes through a fully connected layer with $1$D batch normalization. Subsequently, two blocks are sequentially applied for $2$D transpose convolution, followed by $2$D batch normalization and LeakyReLU activation. The fourth block includes a $2$D convolutional layer with sigmoid activation. In the middle blocks, respectively, $64$ and $16$ filters are utilized for the transpose convolutional layers, with a $3 \times 3$ kernel size and a stride of $2 \times 2$ for upsampling. The final convolution block employs a single filter with a $3 \times 3$ kernel size without any upsampling or downsampling. 

We simultaneously train our multi-encoder multi-decoder architecture for each encoder and decoder, ensuring optimal performance. During training, for static and very-static activities in the presence or absence of disturbers, we only use a human and a human together with a disturber in front of the radar. However, our network generalizes its capabilities to multiple humans in the presence or absence of multiple disturbers. The upper encoder handles the encoding of macro RDIs originating from both static and very-static humans, with and without disturber(s). Similarly, the lower encoder encodes the micro RDIs originating from static and very-static humans, also considering the presence of disturbers. On the other hand, the decoders function as follows, progressing from top to bottom: The first decoder focuses on decoding the latent representation of macro RDIs solely from static humans and static humans in the presence of disturbers. The second decoder decodes the latent representation of macro RDIs exclusively from very-static humans and very-static humans in the presence of disturbers. Conversely, the third decoder is responsible for decoding the latent representation of micro RDIs solely from static humans and static humans with disturbers. Finally, the last decoder decodes the latent representation of micro RDIs originating from very-static humans and very-static humans accompanied by disturbers.

Encoder-decoder pairs are trained specifically for static human(s) with and without disturber(s) by simulating scenarios where a human stands in various locations, exhibiting macro and micro movements in front of the radar. Disturbers such as fans and toy robots are introduced to create realistic interference. The objective of the pairs is to accurately detect static human(s) within the range of 1 to 4 meters, including individuals sitting closely (1 to 2 meters) to the radar. Similarly, additional encoder-decoder pairs are trained for very-static humans with and without disturbers by simulating scenarios where a human sits in different locations, displaying macro and micro movements in front of the radar. These pairs are designed to detect individuals engaging in very-static activities such as sitting or lying down, both in close proximity and at distances ranging from 1 to 4 meters from the radar. Interestingly, even when a person is standing far from the radar (4 to 5 meters), these pairs can still detect their presence. Although standing activity is not categorized as very-static, it generates signals resembling those of very-static individuals due to the weak signal reflection from a standing person positioned far from the radar. The encoder-decoder pairs effectively address a wide range of scenarios, accommodating various edge cases and enabling the detection of human(s) in an activity-free manner. Notably, by focusing on two fundamental human activities—sitting and standing—and incorporating a limited set of disturbers, our solution demonstrates remarkable generalization capabilities across different types of static and very-static human activities involving both micro and macro movements, regardless of the presence or absence of diverse disturbers. Moreover, our solution extends to scenarios involving multiple individuals simultaneously engaging in different static and very-static activities. To approach the presence detection application as an OOD detection problem, we deliberately exclude the use of disturbers in the absence of humans during training. This decision ensures that scenarios lacking human presence remain outside the scope of our training distribution, which is specifically tailored to address OOD detection.

The mean squared error (MSE) function is employed as the loss function in our approach. Since our network consists of four decoders, we aim to minimize four MSEs simultaneously. During the training phase, we utilize the Adamax optimizer \cite{kingma2014adam} and apply various augmentations, including random affine transformations, flips, and scaling. These augmentations not only enhance the robustness of our solution but also contribute to its generalization capabilities. In many radar-based applications, radars are typically mounted at specific heights and tilts. However, thanks to the augmentations incorporated in our approach, our application can adapt to different reasonable heights and tilts. For example, it can operate effectively at TV height without tilt in TV-based applications or be mounted on the corner of a room with a slight tilt for smart home applications.

Therefore, our loss function, as shown in Equation \ref{eq:loss_pres}, aims to minimize the discrepancy between the original data $\textbf{X}^{(i)}$ and its reconstructions obtained through the encoders and decoders in our architecture. The data samples are categorized into four types: macro RDIs of static humans ($M,s$), micro RDIs of static humans ($m,s$), macro RDIs of very-static humans ($M,vs$), and micro RDIs of very-static humans ($m,vs$). The encoders, $E_M$ and $E_m$, are specifically designed to encode the macro RDIs and micro RDIs, respectively. The decoders, $D_{M,s}$, $D_{m,s}$, $D_{M,vs}$, and $D_{m,vs}$ correspondingly decode the different data categories. The MSE loss measures the average squared difference between the original data and the reconstructions over the batch (size of $n$). Since we have four decoders in our architecture, we obtain four MSE values representing the reconstruction losses for each data category. To formulate the overall loss function, we sum the individual MSE losses as follows:
\begin{equation}
\begin{aligned}
\mathcal{L}&  = \frac{1}{n} \sum_{j \in \{m,M\}} \sum_{i=1}^{n} (\textbf{X}^{(i)}_{j,s} - D_{j,s}(E_{j}(\textbf{X}^{(i)}_{j,s})))^2 \\
&+\frac{1}{n} \sum_{j \in \{m,M\}} \sum_{i=1}^{n} (\textbf{X}^{(i)}_{j,vs} - D_{j,vs}(E_{j}(\textbf{X}^{(i)}_{j,vs})))^2\
\end{aligned}
\label{eq:loss_pres}
\
\end{equation}

\subsection{Human and OOD Detection}
In our proposed method, we detect humans and OOD samples simultaneously within a single pipeline. We approach the human detection application as an OOD detection problem. Reconstruction-based methods for OOD sample detection rely on measuring the reconstruction error between the input and the reconstructed output. A sample is classified as OOD if its reconstruction error exceeds a predefined threshold. During the training of HOOD, we exclusively use ID samples that consist of scenarios involving humans and humans with disturbers. Consequently, we expect OOD samples containing only disturbers to exhibit higher reconstruction errors. Leveraging a multi-decoder system, we employ multi-thresholding for analysis. Our network comprises two encoders and four decoders. The encoders receive the macro and micro RDIs of the same input (X) through their respective input gates. The decoders decode the encoded latent codes based on their specific responsibilities. We combine the reconstruction MSEs of macro and micro RDIs separately for the static and very-static human categories. For this combination, we utilize a summation of reconstruction MSEs. Subsequently, we perform multi-thresholding on the combined reconstruction MSEs. If all combined reconstruction errors exceed their respective thresholds, the network classifies the sample as OOD (indicating the absence of humans); otherwise, it is categorized as ID (indicating the presence of humans). To ensure the effective classification of ID data, a threshold is commonly chosen to correctly classify a high percentage (e.g., 95\%) of the ID samples. In the case of HOOD, we set the thresholds to ensure that 90\% of the ID data for each activity (static and very-static) is correctly classified. Following this logic, we define two thresholds corresponding to static and very-static activities.

The pseudocode in Algorithm \ref{alg:pseudo-hood} explains how our human and OOD detector operates in a mathematically rigorous manner.

\begin{algorithm}
\caption{Pseudocode of HOOD inference}
\label{alg:pseudo-hood}
\begin{algorithmic}
\If{\begin{align*}
& \text{MSE}(X_{M,s}, D_{M,s}(E_M(X_{M,s}))) \\
& + \text{MSE}(X_{m,s}, D_{m,s}(E_m(X_{m,s}))) > \text{threshold}_{s} \quad \\ & \textbf{and} \\
& \text{MSE}(X_{M,vs}, D_{M,vs}(E_M(X_{M,vs}))) \\
& + \text{MSE}(X_{m,vs}, D_{m,vs}(E_m(X_{m,vs}))) > \text{threshold}_{vs} 
\end{align*}}
    \State $X \gets \text{No Presence}$
\Else
    \State $X \gets \text{Presence}$
\EndIf
\end{algorithmic}
\end{algorithm}

\section{Experiments}
\label{section:experiment}
In our offline experiments, we utilize a processing unit consisting of an NVIDIA GeForce RTX 3070 GPU, an Intel Core i7-11800H CPU, and a 32GB DDR4 RAM module. We perform our real-time (online) experiments on various CPUs. Additionally, to show the compatibility and feasibility of our pipeline, we extended our tests to include the Raspberry Pi 3B+, which is equipped with an ARM Cortex-A53 CPU and is known for its limited resources. Despite its modest hardware, our model's spatial complexity of 49 MB ensures smooth operation on the Raspberry Pi 3B+. Execution times varied, with the model processing one frame in just 19 ms on the Intel Core i7-11800H CPU, while taking 260 ms on the Raspberry Pi 3B+ with an ARM Cortex-A53 CPU. These findings underscore the real-time capability and adaptability of our model from high-end computing environments to resource-constrained devices.

\begin{table*}[ht]
\centering
\footnotesize
\caption{Main Results. HOOD performance comparison with other popular methods. All values are shown in percentages except time $\uparrow$ indicates higher values are better, while $\downarrow$ indicates lower values are better.}
\label{Results}
\setlength\tabcolsep{0pt}
\begin{tabular*}{\textwidth}{@{\extracolsep{\fill}}c|cccccccccc}
    \toprule
    &
    & \multicolumn{4}{c}{Static} &\multicolumn{4}{c}{Very-Static} &{Test Time }\\
    \cmidrule{3-6} \cmidrule{7-10}
    Architecture & Method 
    & AUROC
    & AUPR\textsubscript{IN}
    & AUPR\textsubscript{OUT}
    & FPR95 
    & AUROC 
    & AUPR\textsubscript{IN}
    & AUPR\textsubscript{OUT}
    & FPR95
    & (seconds) \\
    &  
    &  $\uparrow$
    &  $\uparrow$
    &  $\uparrow$
    &  $\downarrow$
    &  $\uparrow$
    &  $\uparrow$
    &  $\uparrow$
    &  $\downarrow$
    &  $\downarrow$ \\
    \midrule
    & MSP \cite{b1} & 53.82&60.27&40.72&97.51&56.75&65.71&41.54&91.31&298 \\
    & ODIN \cite{b2} &88.67&92.81&80.77&62.08& 81.44&89.25&67.20&73.81& 1209\\
    & ENERGY \cite{b7} &47.01&57.09&38.11&95.34& 54.55&65.45&45.37&83.22&288 \\
    RESNET-34 \cite{resnet}& MAHA \cite{b4} &37.97&49.84&50.17&73.83 & 54.95&67.08&52.48&73.81&3599 \\
    &FSSD \cite{b6} &90.77&92.01&88.00&44.44&80.77&88.26&70.92&65.10&3784 \\
    &OE\cite{b8} &64.75&66.27&63.67&72.36&77.96&84.88&69.71&68.65&289 \\
    &GRADNORM\cite{b14} &53.38&59.02&41.56&96.66&54.01&63.68&39.72&91.79&878 \\
 &REACT\cite{b28} &41.67&54.17&33.54 & 97.76&53.46&64.99&42.55&86.09&297\\
    &MAXLOGIT\cite{hendrycks2022scaling} &50.17&59.64&39.34&94.98&51.59&65.84&34.35&94.85&297 \\
    &KL\cite{hendrycks2022scaling} &52.95&62.89&45.40&90.61&64.65&75.76&49.86&84.83&287 \\
    \midrule
    \midrule
    \addlinespace
    
    & MSP\cite{b1} &19.33&43.63&25.57&100&30.43&53.32&24.25&99.50&578 \\
    & ODIN \cite{b2} &43.21&55.57&34.56&96.97&49.64&64.99&35.46&92.75&2466 \\
    & ENERGY\cite{b7} &50.52&57.03&41.60&93.75&48.51&62.02&33.70&94.69&600 \\
    RESNET-50 \cite{resnet}& MAHA \cite{b4} &82.29&87.20&78.13&60.16&81.25&89.77&70.48&66.50&7249 \\
    &FSSD\cite{b6} &85.57&90.43&77.41&71.29& 72.85& 85.59& 57.15&82.42& 8041\\
    &OE\cite{b8} &78.00& 83.82& 71.15&70.81& 77.81& 86.09&67.06 &69.01& 581 \\
    &GRADNORM\cite{b14} &56.11& 60.96& 50.86&81.52& 40.41& 57.46& 30.96&93.22& 1850 \\
    &REACT\cite{b28}  &63.59&69.85&49.44&94.70& 54.07& 68.97& 37.30&92.27& 620\\
    &MAXLOGIT\cite{hendrycks2022scaling} &49.93& 60.66& 39.17&94.86& 46.10& 65.00& 29.69&98.53& 642 \\
    &KL\cite{hendrycks2022scaling} &66.65& 75.45 &53.49&90.18&29.68& 54.70& 23.81& 99.23& 573 \\
    \addlinespace
    \midrule
    \midrule
    & MSP\cite{b1} &62.53&68.30&57.66&80.07 & 54.20& 67.67 &41.62 &87.32 &959  \\
    & ODIN \cite{b2} &76.43 &78.63& 74.19&60.05& 65.78& 74.43& 54.05&81.46& 3955\\
    & ENERGY \cite{b7} &71.49 &80.13& 58.72 &85.44& 53.99 &72.14& 35.85 &94.43& 938 \\
    \centering
    RESNET-101 \cite{resnet}& MAHA \cite{b4} &70.69& 75.85 &69.24 &64.07&83.17& 90.49& 73.94 &61.86& 10282 \\
    &FSSD\cite{b6} &86.07& 91.09& 79.21&66.48& 74.28& 86.62&61.25&75.15& 29501 \\
    &OE\cite{b8} &73.90& 79.06& 69.05&68.63& 76.12& 83.24& 66.17&67.88& 932 \\
    &GRADNORM\cite{b14} &52.02& 59.95& 44.98&90.20&46.56& 63.31& 32.76&94.10& 7472 \\
    &REACT\cite{b28}  &41.25& 55.09& 35.95&94.64& 38.63& 57.11& 31.89& 93.21& 988 \\
    &MAXLOGIT\cite{hendrycks2022scaling}  &50.65& 59.21&39.63&94.95&50.11& 64.12 &34.04& 95.12 & 1049 \\
    &KL\cite{hendrycks2022scaling} &71.41& 75.62& 68.17& 67.35& 61.82& 69.25& 56.94& 72.71& 939 \\
\addlinespace
    \midrule
    \midrule
    
Ours &HOOD &\textbf{95.71}&\textbf{97.09}&\textbf{93.80}&\textbf{23.67}&\textbf{93.02}&\textbf{96.11}&\textbf{87.70}&\textbf{34.46}&\textbf{67} \\
    
    \bottomrule
\end{tabular*}
\end{table*}

\begin{table*}[ht]
\centering
\footnotesize
\caption{Ablation study on the impact of E-RESPD on the performance of HOOD. All values are shown in percentages. $\uparrow$ indicates that higher values are better, while $\downarrow$ indicates that lower values are better.}
\label{ablation}
\setlength\tabcolsep{0pt}
\begin{tabular*}{\textwidth}{@{\extracolsep{\fill}}c|cccccccc}
    \toprule
     
    & \mc{4}{\mytab{Static}} & \mc{4}{\mytab{Very-Static}}
    \\
  
    \cmidrule{2-5} \cmidrule{6-9} 
\centering
      
    & AUROC
    & AUPR\textsubscript{IN}
    & AUPR\textsubscript{OUT}
    & FPR95 
    & AUROC 
    & AUPR\textsubscript{IN}
    & AUPR\textsubscript{OUT}
    & FPR95  
      \\
    &  $\uparrow$
    &  $\uparrow$
    &  $\uparrow$
    &  $\downarrow$
    &  $\uparrow$
    &  $\uparrow$
    &  $\uparrow$
    &  $\downarrow$
      \\  
      
    \midrule
 HOOD w/o E-RESPD&85.14&88.50&80.73&55.12&76.07&85.91&61.68&76.05   \\     
HOOD&\textbf{95.71}&\textbf{97.09}&\textbf{93.80}&\textbf{23.67}&\textbf{93.02}&\textbf{96.11}&\textbf{87.70}&\textbf{34.46}\\

    \addlinespace

    
    \bottomrule

\end{tabular*}
\end{table*}

\subsection{Dataset and Evaluation Metrics}

In this study, we record our own data utilizing Infineon's BGT60TR13C $\SI{60}{\giga\hertz}$ FMCW radar sensor with four individuals in various indoor places, including 28 rooms from houses, classrooms, and offices. The radar sensor is placed at a height of around two meters and tilted 15 degrees from the front side.
We divide data from all rooms into separate sets for training and testing purposes. Specifically, 14 rooms are allocated for the training, and the remaining 14 rooms are used for the testing. The dataset consisted of two types of samples: ID and OOD. OOD samples consist of common moving household objects such as table fans, stand fans, a remote-controlled (RC) toy car, a robot vacuum cleaner, etc. In addition, we divide the ID samples into two categories: static and very-static. The static scenarios involve a person standing alone or standing with a disturber object present in the scene. On the other hand, the very-static scenarios consist of a person either sitting or lying alone, as well as with a disturber object present in the scene. Throughout the recordings, both the disturber objects and individuals engage in their activities at distances ranging from 1 to 4 meters away from the radar.

During training, we use ID samples from static and very-static categories of sitting and standing activities in the presence or absence of disturbers. The disturbers in the training set include a stand fan, a table fan, an RC toy robot, and moving lamps. For testing, we use both ID and OOD samples and expand the variety of disturber types, including an RC toy car, boiling water from a kettle, running water from a tap, a vacuum cleaner, a smart vacuum cleaner, swinging laundry, a stand fan, a table fan, and a flying paper bag. Additionally, the very-static category of ID samples is increased by including the scenario of a person lying down.

We build the training set using only ID samples, consisting of 224063 frames from the static category and 226484 frames from the very-static category. For the test set, we use 108952 frames from the static category, 137325 frames from the very-static category as ID frames, and 73960 OOD frames.

Our study utilizes several common evaluation metrics in the OOD detection field: AUROC, AUIN, AUOUT, and FPR95. \textbf{AUROC} is the measure of the area under the receiver operating characteristic (ROC) curve. \textbf{AUPR\textsubscript{IN}} is the area under the precision-recall curve when considering ID samples as positives.     \textbf{AUPR\textsubscript{OUT}}  is the area under the precision-recall curve when considering OOD samples as positives. \textbf{FPR95}  corresponds to the false positive rate (FPR) when the true positive rate (TPR) is at 95\%. On the other hand, the \textbf{Test Time} represents the inference time in seconds required to evaluate all test samples. It also shows the minimal time complexity of our model.

 We compare the performance of HOOD with 10 SOTA methods using the above-mentioned common OOD detection metrics. To ensure a comprehensive comparison, we separately train ResNet34, ResNet50, and ResNet101 backbones\cite{resnet} with our static and very-static ID categories in a two-class classification manner. As the nature of the OOD detection tasks, the backbones are never fed with the OOD data during training. Subsequently, we utilize the pre-trained models to employ the SOTA methods and evaluate the metrics. The superiority of HOOD over the SOTA approaches is presented in Table \ref{Results}. For comparisons with the SOTA methods, we use the identical train and test datasets employed in the HOOD framework. Furthermore, we perform an ablation study to assess the impact of incorporating E-RESPD into HOOD. Table \ref{ablation} shows the significant effect of E-RESPD in enhancing HOOD's performance.

 \subsection{Conducted Experiments}
We perform a series of real-time experiments with different individuals to demonstrate the performance of HOOD in different scenarios. By doing so, we want to quantify its generalization performance in various cases, including challenging scenarios and for different individuals that are not observed by the model during training. For the experimental setup, we position the radar at different heights ranging from 1 meter to 2.5 meters and adjust the tilt degree from 0 to 30 degrees. The distances and angles of individuals and disturber objects with respect to the radar vary. They may be located within the same or different range and angle relative to the radar. The specific details of the scenarios are as follows:

\begin{enumerate}
\item \textbf{Normal-Cluttered Room:}
This experiment focuses on indoor rooms such as houses, offices, and classrooms where disturber objects are almost stationary, and they characterize insignificant disruption within the room. We refer to such indoor places as normal-cluttered rooms. Importantly, there are no individuals present in the room during this case. We evaluate the performance of the HOOD system in normal-cluttered environments to assign \textit{no presence}.

\item \textbf{Heavy-Cluttered Room:}
In this experiment, we examine scenarios where there are no individuals present in the room, but in contrast to normal-cluttered rooms, disturber objects cause significant movements. Commonly found disturbers in indoor environments include stand fans, table fans, RC toy cars, and other similar objects. This experiment aims to assess HOOD's capability to assign \textit{no presence} accurately.

\item \textbf{Human Presence in Normal-Cluttered Room:}
We investigate scenarios where a human engages in sitting or standing activities in a normal-cluttered room. Importantly, these scenarios are not limited to a single person; we also examine multiple-person scenarios. These involve cases where some people are sitting while others are standing, as well as situations where all individuals are sitting or standing simultaneously. Furthermore, in the additional tests, we also test individuals sitting or standing with their backs facing the radar. This experiment aims to assess HOOD's capability to assign \textit{presence} considering various postures and multiple-person configurations correctly.

\item \textbf{Human Presence in Heavy-Cluttered Room:}
We focus on scenarios where individuals can be either sitting or standing in a heavy-cluttered room. The experiment includes various combinations of sitting and standing activities, providing diverse challenges for the system. The goal is correctly assigning \textit{presence} to complex scenarios involving multiple individuals, different postures, and disturbers.

\item \textbf{Partially Visible Human in Normal \& Heavy-Cluttered Room:} In this scenario, an individual is standing or sitting in a normal or heavy-cluttered room. However, a table may partially block the human body, or the human stays at an angle at which only some body parts are in the radar's field of view. Our system aims to accurately detect the \textit{presence} of a person, even when their body is partially visible or obscured by a table or other objects in the room.

\item \textbf{Edge Case Analysis:}
We also test various edge cases to determine the system's performance boundaries. For example, we test a human standing and holding their breath, a human sitting and holding their breath, and a person lying down and crouching, resulting in partial obscuration of the chest area. Additionally, we place a corner reflector alongside individuals who are sitting or standing. We evaluate the system's ability to assign \textit{presence} through these tests by handling these challenging scenarios accurately.
\end{enumerate}

For additional experimental scenarios and a comprehensive demonstration of HOOD, including detailed demo videos, please visit the project page\footnote{\url{https://muskahya.github.io/HOOD}}.

\section{Conclusion}
\label{section:conclusion}
This study presents HOOD, a real-time and robust human presence and OOD detection method specifically designed for indoor environments using low-cost $\SI{60}{\giga\hertz}$  FMCW radar. By exploiting a reconstruction-based architecture and leveraging radar macro and micro RDIs, HOOD effectively detects the presence of humans in the presence or absence of moving and stationary disturbers. As an OOD detector, it accurately identifies and classifies disturbers as OOD when no humans are detected, providing reliable predictions for different human scenarios. The evaluations conducted on a dataset collected by the $\SI{60}{\giga\hertz}$ FMCW radar demonstrate the superior performance of HOOD, achieving high AUROC and low FPR95 values. Comparisons with SOTA OOD detection methods further confirm the effectiveness of HOOD. Moreover, the real-time capability of HOOD enhances its practical utility for various indoor applications, and its compatibility with Raspberry Pi 3B+ underscores its versatility across different hardware environments. Overall, HOOD presents a promising real-time human presence and OOD detection solution with exceptional generalization capabilities. A potential extension for the practical implementation of HOOD involves integrating it into a microcontroller, enabling autonomous operation in conjunction with the radar system. By embedding HOOD directly onto a microcontroller, it becomes possible to create a self-contained and independent solution for human presence and OOD detection. 

\end{document}